\documentclass{article}
\pdfoutput=1
\usepackage{jinstpub} 
\usepackage{graphicx}
\usepackage{here}
\usepackage{amssymb}
\usepackage{amsmath}
\usepackage{subfigure}
\usepackage{lineno}
\usepackage{enumerate}
\usepackage{lineno}
\usepackage{hyperref}
\usepackage{units}
\usepackage{xcolor}

\title{\boldmath A FLUKA study towards predicting\\hadron-specific damage due to high-energy hadrons\\ in inorganic crystals for calorimetry}

\author[]{G. Dissertori,}
\author[1]{C. Martin Perez\note{ now at Laboratoire Leprince-Ringuet, Ecole Polytechnique, Palaiseau, France},}
\author[2]{F. Nessi-Tedaldi\note{ corresponding author}}

\affiliation[]{ETH Zurich, Institute for Particle Physics and Astrophysics\\ 8093 Zurich, Switzerland}

\emailAdd{Francesca.Nessi-Tedaldi@cern.ch}

\abstract{Hadrons emerging from high-energy collisions, as it is the case for protons and pions at the CERN Large Hadron Collider, can produce a damage to inorganic crystals that is 
specific and cumulative. The mechanism is well understood as due to bulk damage from fragments caused by fission. In this paper, the existing experimental evidence for lead tungstate, LYSO and cerium fluoride is summarised, a study using FLUKA simulations is described and its results are discussed and compared to measurements. The outcome corroborates the confidence in the predictive power of such simulations applied to inorganic scintillators, which are relevant to their adoption as scintillators for calorimetry.
}

\begin{document}
\maketitle
\flushbottom

\section{Introduction}
The exposure of inorganic calorimeter crystals to large fluences of high-energy hadrons became an issue needing investigation with the construction of the Large Hadron Collider (LHC) and of its detectors at CERN. The choice of lead tungstate (${\mathrm{PbWO}}_4$, or simply PWO) crystals for the CMS electromagnetic calorimeter (ECAL), where non-negligible exposures to high-energy hadrons were expected~\cite{r-ECALTDR}, raised questions on a possible, hadron-specific damage component, while studies of the ionising damage and its minimisation were already well advanced through crystal quality optimisation~\cite{r-ECALRAD}.

A series of pioneering systematic irradiation studies allowed to single out and understand the mechanism of a hadron-specific damage component. Through exposure to 20 -  24 GeV protons, it was shown that lead tungstate crystals' light transmission is affected in a cumulative way, with no spontaneous recovery at room temperature~\cite{r-LTNIM}. Further, the scintillation mechanism was proven to be left unaffected~\cite{r-LYNIM}. Complementary irradiations with charged pions~\cite{r-PIONNIM} allowed to understand the scaling behaviour between different particle types and the predictive power achievable within the same crystalline material through simulations using the FLUKA package~\cite{r-FLU}. In parallel, it became possible to draw predictions from data taken {\em in situ} for the CMS ECAL detector longevity~\cite{r-CMSlongevity} which led to the decision to replace its endcaps in view of the LHC upgrade for high-luminosity running (HL-LHC)~\cite{r-HL-LHC}. The predictions have recently been confirmed by observations of signal losses in the CMS ECAL that are by now quite significant, especially in the most exposed, high-$\eta$ regions of the detector~\cite{r-ECALLOSS}.

The need to find scintillating materials suitable for calorimetry in view of the HL-LHC upgrade led to study hadron damage in cerium-doped lutetium-yttrium orthosilicate (LYSO)~\cite{r-LYSONIM} and in cerium fluoride  (${\mathrm{CeF}}_3$)~\cite{r-CEFNIM}. While LYSO exhibits a cumulative damage, although of smaller amplitude than lead tungstate, cerium fluoride spontaneously recovers from damage at room temperature, in a way which is consistent with a purely ionising damage mechanism~\cite{r-AUFCEF}.

Through all the studies above, the understanding was reached that a hadron-specific damage mechanism is at work in PWO and LYSO, an effect which was also observed in BGO~\cite{r-BGOHADDAM,r-BGOFN}. It is due to the extremely high energy loss of fission fragments along their short path in the crystalline matter, leaving small regions of damage behind, that act as scatterers for the scintillation light. The effective reduction of light output is macroscopically observed as a loss in light transmission. Indirect evidence for it was the Rayleigh scattering $\left (\lambda^{-4}\right )$ behaviour~\cite{r-RS} of the damage amplitudes observed in \cite{r-LTNIM}, the visualisation of scattering centres by shining laser light through irradiated PWO and LYSO crystals and the fact that the scattered light is polarised~\cite{r-FISSNIM,r-LYSONIM}.
All of these effects were not observed in cerium fluoride~\cite{r-CEFNIM}, as expected, since the material consists of light elements that do not undergo fission~\cite{r-FISS}.
An ultimate proof was reached through the direct visualisation of fission tracks from proton irradiation of lead tungstate, following a method developed for geochronology and used for mineral dating~\cite{r-FISSNIM}. 

The damage regions that cause hadron-specific signal losses by scattering the scintillation light have dimensions that are driven by the fission fragments' path lengths. In turn, these are linked to the fragments' stopping power. Fragment production and energy losses are processes that lie in the realm of FLUKA simulations. The possibility to verify how well FLUKA can reproduce measured quantities and thus to validate its predictive power for different materials was therefore found to be of interest for an informed design of future calorimeters and has thus inspired the work presented herein.

\section{Fission track formation}
\label{s-FISS}
The regions of damage left by fission fragments in minerals are commonly called ``fission tracks'' in literature. They have been studied foremost as a natural phenomenon occurring in Muscovite mica that is found near an uranium-containing ore and in meteorites that have been exposed to high-energy hadrons from outer space. The understanding of the fission track formation mechanism has been reached through a vast effort documented in literature~\cite{r-FLE}. The track formation is explained in \cite{r-FLEME}  and references therein as a 3-stage process where a) a charged particle passage causes ionisation, b) ions are ejected from their lattice locations due to Coulomb repulsion and c) a region of atomic disorder is left behind. The main parameter involved is the primary ionisation density, i.e. the number of electrons displaced per unit length.
\begin{figure}[thb]
\begin{center}
 \begin{tabular}[h]{cc}
{\mbox{\includegraphics[width=80mm]{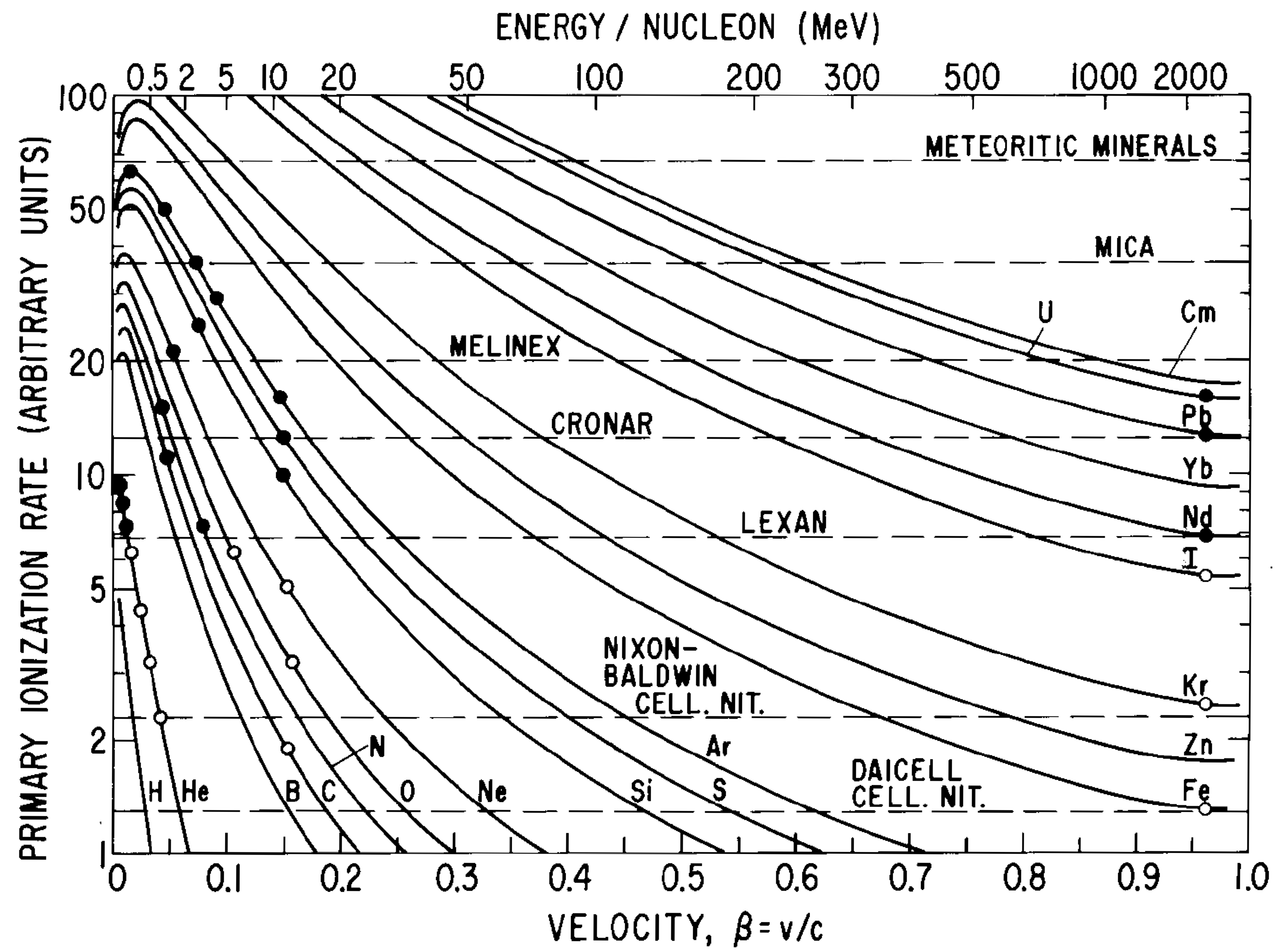}}} &
{\mbox{\includegraphics[width=60mm]{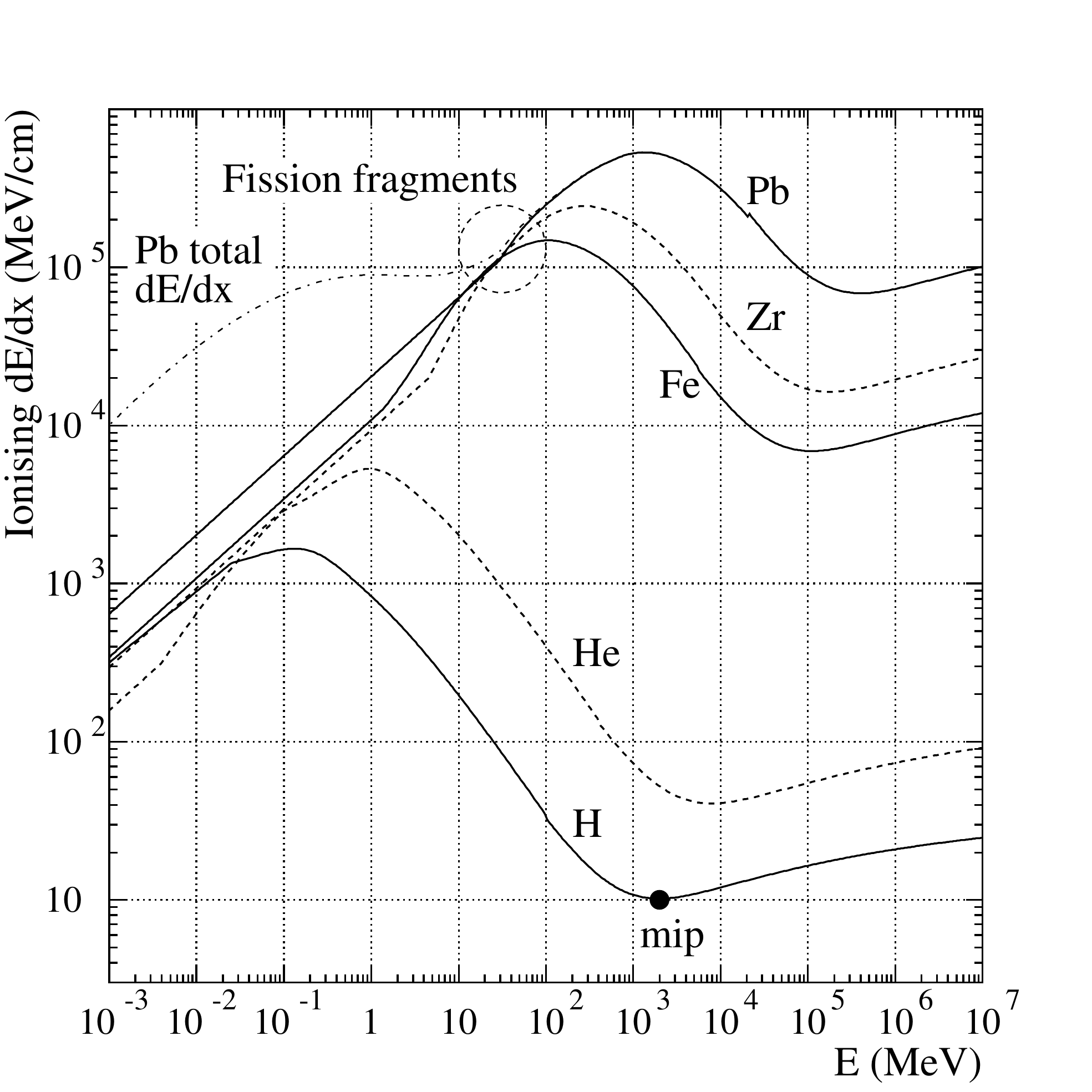}}}
   \end{tabular}
\end{center}
\caption{Left: Primary ionisation in various nonconducting solids as a function of velocity and of energy per nucleon for various projectile nuclei (figure 2 from~\cite{r-FLE} ). The experimental points for different ions are given as open circles for no tracks observed, and as filled circles if tracks are observed. The dashed horizontal lines indicate the thresholds for track recording. Right: Simulated ionising energy loss for different fragments in lead tungstate (figure 1 from ~\cite{r-LTNIM}).  The dashed circle shows the typical dE/dx-values of fission fragments at the beginning of their track. }
\label{f-dEdx}
\end{figure}
Existing data are collected in figure \ref{f-dEdx} (left) from \cite{r-FLE}. It is evident that track formation is observed above a material-specific ionisation density threshold which is common for all projectile hadrons in a given material, with inorganic materials exhibiting higher thresholds than organic ones. In inorganic crystals, tracks are observed to have been formed by projectiles whose mass number is larger than 30, while no tracks are formed for $A < 20$.
The ionisation rate in figure \ref{f-dEdx} (left) is given in arbitrary units, however the energy loss per unit distance (dE/dx) threshold value for producing tracks is found by comparison with the plot in figure \ref{f-dEdx} (right), from \cite{r-LTNIM}.
For heavy nuclei, track formation occurs for dE/dx $\gtrsim1\times 10^5$ MeV/cm.
\begin{figure}[b]
\begin{center}
 \begin{tabular}[h]{cc}
{\mbox{\includegraphics[width=55mm]{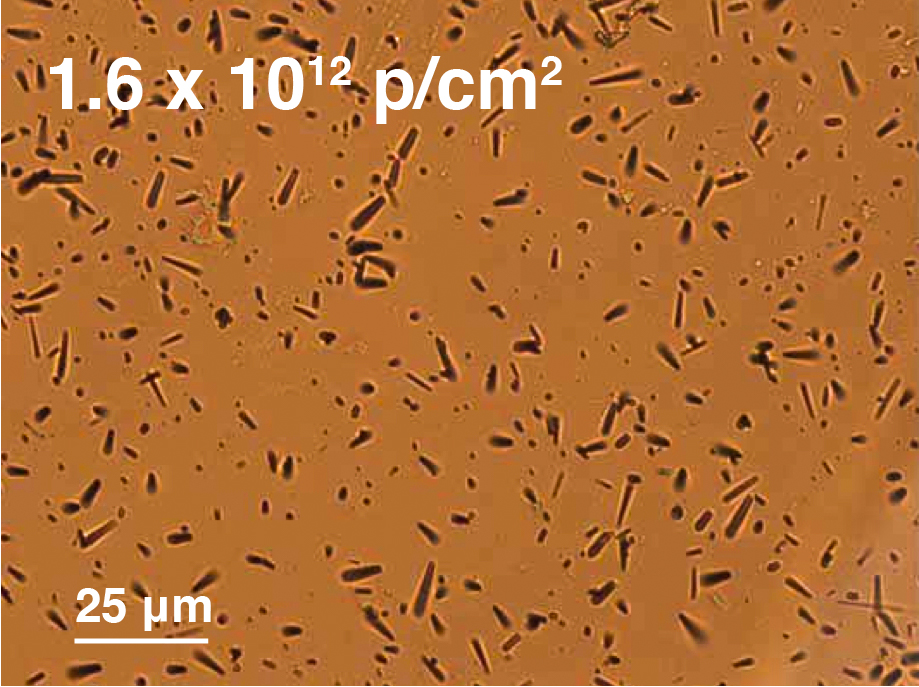}}} &
{\mbox{\includegraphics[
bb = -10 55 700 400, clip, 
width=85mm]{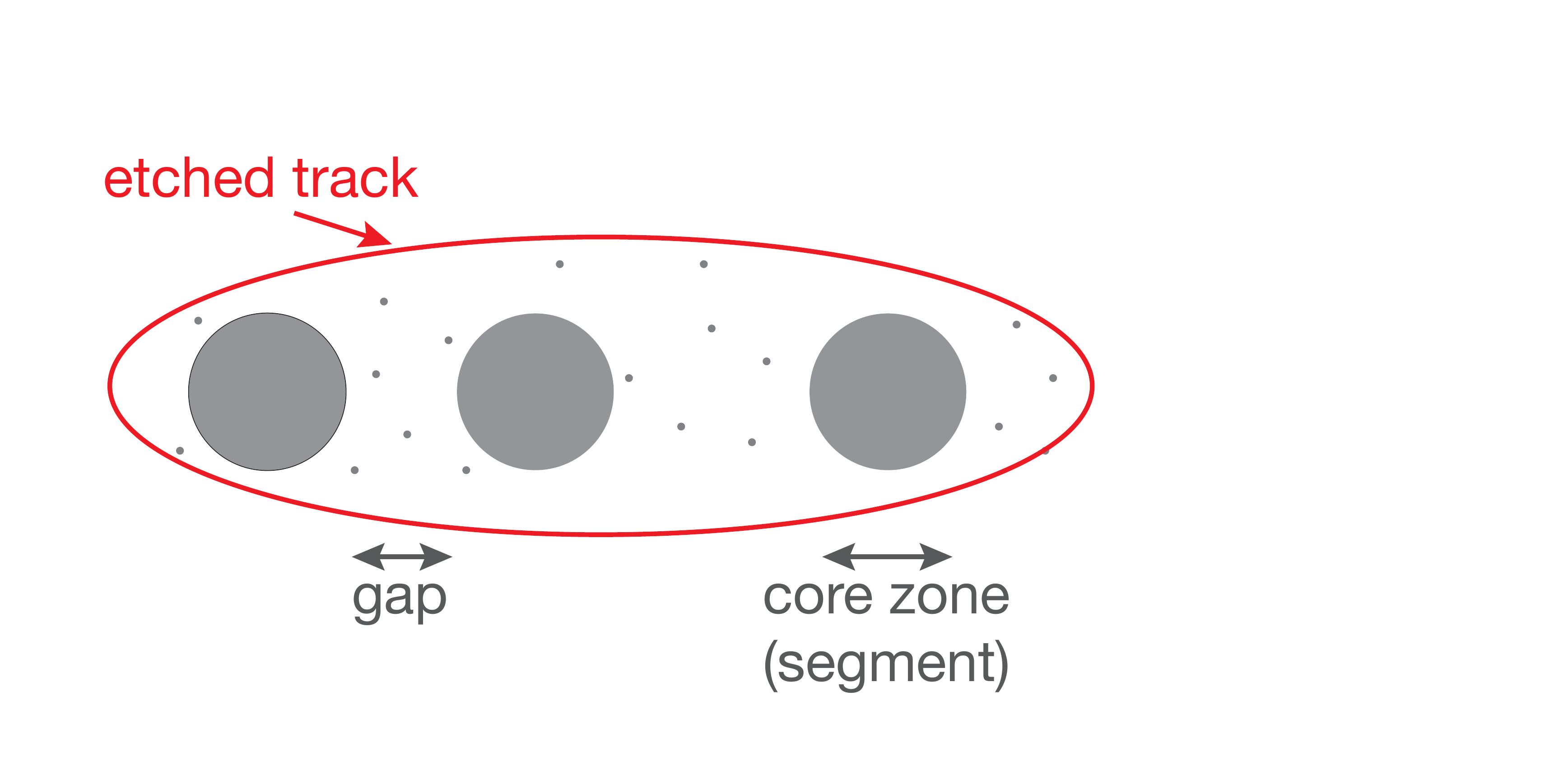}
}}
   \end{tabular}
\end{center}
\caption{Left: Figure from transmitted-light optical-microscopy of etched fission tracks from lead tungstate recorded in a mica slide.
Right: Schematic structure of a fission track.\label{f-PWOTRK}}
\end{figure}

Once tracks have been formed, light scatters against them. This phenomenon has been visualised for lead tungstate in~\cite{r-FISSNIM},  where tracks have been revealed through a technique used in geochronology: tracks crossing a surface are revealed through etching, since their material is in a disordered, ``metamict'' state~\cite{r-FLEME,r-WAG}, which is more easily soluble than the surrounding crystal lattice. An image of fission tracks for lead tungstate is presented in figure \ref{f-PWOTRK}, which has been obtained as described in~\cite{r-FISSNIM}, and where typical etched track lengths of a few~$\mu$m are evident.

One needs to make a distinction, however, between tracks that have been etched for visualisation purposes, and latent tracks, which are those affecting crystals' light transmission in calorimeters. Latent track studies tell us that 
\begin{enumerate}
\item tracks can be composed by regions of extended defects (``core zones'' in literature, called segments herein, see figure \ref{f-PWOTRK} right), with gaps of sparse damage in between~\cite{r-DAR, r-CHA};
\item segment diameters amount to a few nanometers~\cite{r-YAD} and their length depends on the projectile;
\item light scatters against such segments;
\item the gaps in between segments contain only point defects and are etched more slowly~\cite{r-WAG}.
\end{enumerate}

\section{Rayleigh Scattering}
\label{s-RS}
Having observed that fission tracks in scintillating crystals cause scattering of the scintillation light that exhibits all the features of Rayleigh scattering (RS), we summarize herein the relevant quantities of the RS approximation. We shall then attempt obtaining a theoretical input to the equations from FLUKA.

The cross section for Rayleigh scattering, $\sigma_{RS}$, is given by
\begin{equation}
\sigma_{RS} = \frac{8\pi}{3} d^{\,6}\left(\frac{2\pi n_{med}}{\lambda_0}\right)^{4}\left( \frac{m^2-1}{m^2+2}\right)^2
\label{e-sRS}
\end{equation}
with {\em d} the dimension of the scatterers,
$\lambda_0$ the vacuum wavelength of the scintillation light and $n_{med}$ the refractive index of the crystal), and finally $m=n_{sc}/n_{med}$ the ratio of the refractive index of the scatterers to that of the crystal ~\cite{r-COX}.

It should be recalled that the RS approximation is applicable for spherical particles that have radii small compared to the wavelength of the scattered light.
With the so-called "size parameter" $x$ defined~\cite{r-COX} as
\begin{equation}
x = \frac{2\pi n_{med} d}{\lambda_0}
\label{e-RSx}
\end{equation}
the requirement is usually formulated as
\begin{equation}
x m \ll 1 .
\label{e-RSr}
\end{equation}

The fraction $F_{RS}$ of light that gets Rayleigh-scattered when going through a crystal of length ${\cal L}$ is given by
\begin{equation}
F_{RS}\; = \; N \times {\cal L} \times\sigma_{RS} 
\label{e-FRS1}
\end{equation}
where {\em N} is the density of scatterers.

The intensity $I$ of light transmitted through the crystal is to first order
\begin{equation}
I = I_0 e^{-\mu {\cal L}} \simeq  I_0 \left( 1 - \mu {\cal L}\right)
\label{e-I}
\end{equation}
for $\mu < {\cal L}$,
with $I_0$ the transmitted light intensity before the creation of scatterers through irradiation. By combining equations~\ref{e-FRS1} and \ref{e-I}, one can express $F_{RS}$ as
\begin{equation}
F_{RS}\; = \; \frac{I - I_0}{I_0} \simeq \mu {\cal L}.
\label{e-FRS2}
\end{equation}

By combining eqs.~\ref{e-sRS}, ~\ref{e-FRS1} and ~\ref{e-FRS2} one obtains for the ratio $R_{\mu}$ of the induced absorption coefficients between PWO and LYSO:
\begin{equation}
R_{\mu} = \frac{\mu_{PWO}}{\mu_{LYSO}} =  \left( \frac{d_{PWO}}{d_{LYSO}} \right)^6\left(\frac{N_{PWO}}{N_{LYSO}}\right)\left(\frac{\lambda^{LYSO}}{\lambda^{PWO}}\right)^4\left(\frac{\left( \frac{m_{PWO}^2-1}{m_{PWO}^2+2}\right)}{\left( \frac{m_{LYSO}^2-1}{m_{LYSO}^2+2}\right)}\right)^2.
\label{e-RRS}
\end{equation}
The emission spectrum~\cite{r-LAM} and the index of refraction as a function of wavelength for LYSO~\cite{r-nLYSO} and for PWO~\cite{r-nCHI} have been used to obtain the average scintillation light wavelength in each material. While the scintillation light wavelength in vacuum is nearly identical for LYSO and PWO, in the material it amounts in average to $\lambda^{LYSO}=248$ nm and $\lambda^{PWO}=198$~nm because of the different index of refraction of the two crystals, with uncertainties discussed in section~\ref{s-RSratio}.
Measurements of the ratio between the refraction index of a metamict state and the one of the crystalline lattice, $m$, for different crystals lie around $0.93 \pm 0.01$~\cite{r-QUA,r-ZIR}.
We thus use $m_{PWO}=m_{LYSO}=0.93$ and propagate the uncertainty on each one of them into the determination of the uncertainty on $R_{\mu} $ (see section~\ref{s-RSratio}) that we call $\Delta R_\mu$. With the values above, one obtains
\begin{equation}
R_{\mu} = \frac{\mu_{PWO}}{\mu_{LYSO}} = 2.46 \times  \frac{N_{PWO}}{N_{LYSO}}  \times \left( \frac{d_{PWO}}{d_{LYSO}} \right)^6
\label{e-RshortS}
\end{equation}
The density of scatterers and their dimension need to be determined in FLUKA, to allow comparing with the measured ratio, $R_{\mu} = 4.5 \pm 0.2$\footnote{uncertainty inferred from the publication}~\cite{r-LYSONIM}.
\section{FLUKA simulation setup}
\label{s-FLU}
For the simulations study, which involves particle interactions and transport in matter, the FLUKA package~\cite{r-FLU} has been adopted. 
FLUKA offers the advantage of focusing on the general particle interactions and their consequences, it is especially designed for the evaluation of radiation environments, it allows for averaged material compositions and simple combinatorial geometry and it enables accessing several quantities related to radiation damage. Even non-standard, additional information can be retrieved with user-written routines, as it has been the case for the study reported herein. The results have been obtained with FLUKA version 2011.2x.3 for a total of one million generated primaries reaching each crystal front face.

The experimental setups of the irradiations and measurements~\cite{r-LTNIM,r-LYSONIM,r-CEFNIM} have been reproduced in FLUKA. The main parameters of the irradiations were the geometry and composition of the exposed crystals and of the surrounding elements of the irradiation facility IRRAD1 located in the CERN PS T7 beam line~\cite{r-IR1}, the proton beam energy of 20 GeV, respectively 24 GeV, and the beam spot distribution.
The crystals studied are:
\begin{enumerate}
\item a  lead tungstate crystal in the shape of a truncated equal-sided pyramid, 23 cm long, with a minor face of
$22 \times 22$ mm$^2$ and a major face of $25 \times 25$ mm$^2$ yielding a total volume of 127 cm$^3$~\cite{r-LTNIM};
\item a LYSO parallelepiped with dimensions $25\times 25 \times100$ mm~$^3$, cerium-doped, with chemical formula Ce$_{0.003}($Lu$_{0.9}$Y$_{0.1})_{1.997}$SiO$_5$~\cite{r-LYSONIM};
\item a cerium fluoride parallelepiped with dimensions $21\times16 \times141$ mm~$^3$, Barium-doped at a 0.2\% level, with formula Ba:CeF$_3$~\cite{r-CEFNIM}.
\end{enumerate}
The determination of quantities of interest inside the crystals have been performed according to the experimental measurement conditions. In particular, the light transmission, wherefrom the radiation-induced absorption coefficient is determined, was measured in the experiments with a
spectrophotometer whose light beam was aligned on axis with the crystal. Its light beam spot size was 7 mm wide and 10 mm high~\cite{r-LTNIM}.
\section{FLUKA simulation results}
\subsection{Simulations at the infinitesimal level}
\label{s-INF}
As a first approach, the main quantities characterising the radiation field have been explored (density of inelastic interactions, dose, density of charged hadrons and of neutrons, displacement-per-atom) as they had been successful in reproducing the scaling of damage in lead tungstate between different particle types and energies~\cite{r-PIONNIM}. However, when trying to scale between different materials, the ratios of the quantities above, which are of no fundamental nature, are insufficient to reproduce observations. A ``microscopic'' approach has thus been adopted in the study presented here, where particles are followed at every step of their propagation in matter, and where regions of damage are reconstructed following the experimental observations described in section~\ref{s-FISS}.

\begin{figure}[t]
\begin{center} 
 \begin{tabular}[h]{ccc}
{\mbox{\includegraphics[viewport = 260 200 780 560, clip, width=50mm]{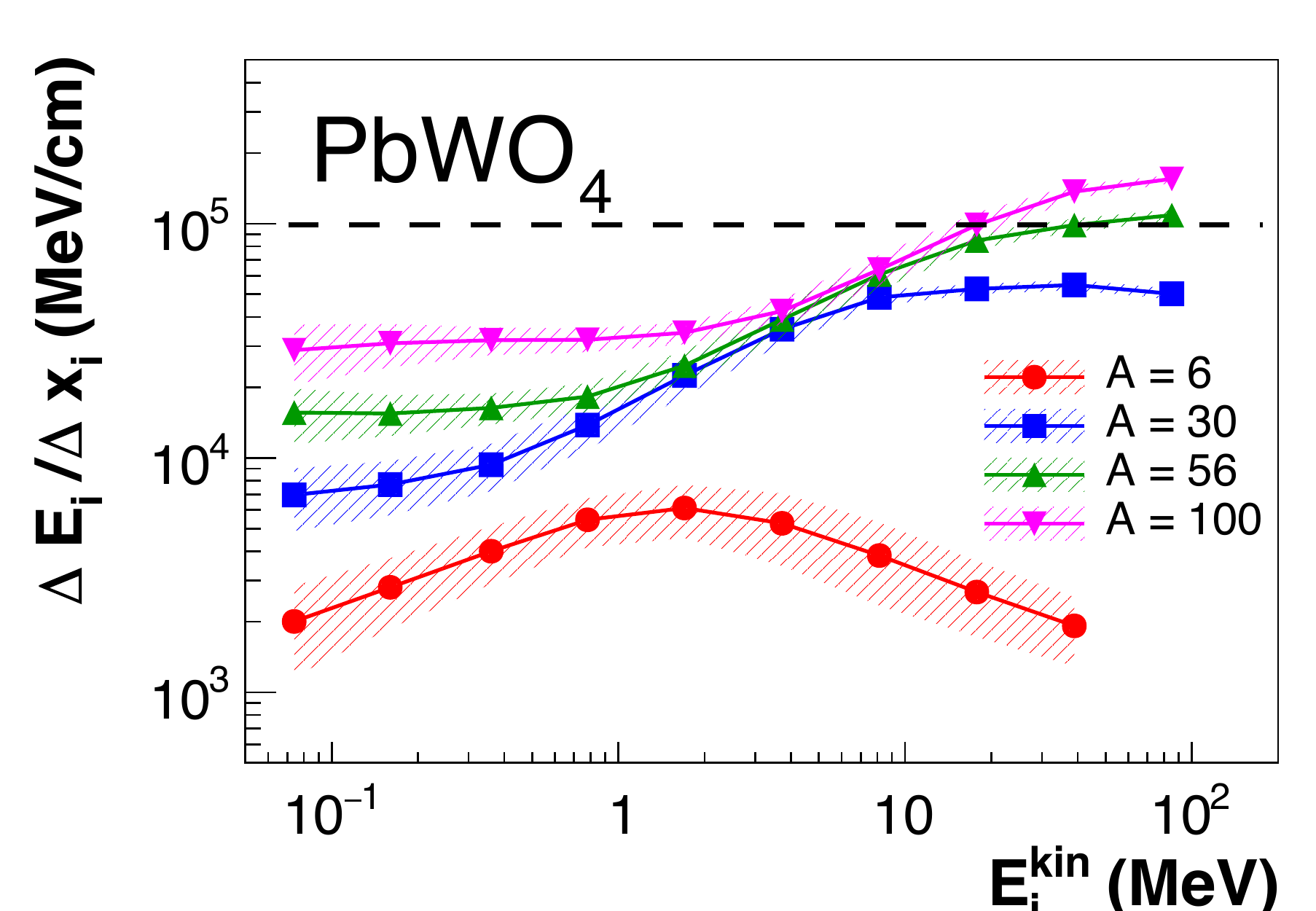}}} &
{\mbox{\includegraphics[viewport = 25 200 555 580, clip, width=46mm]{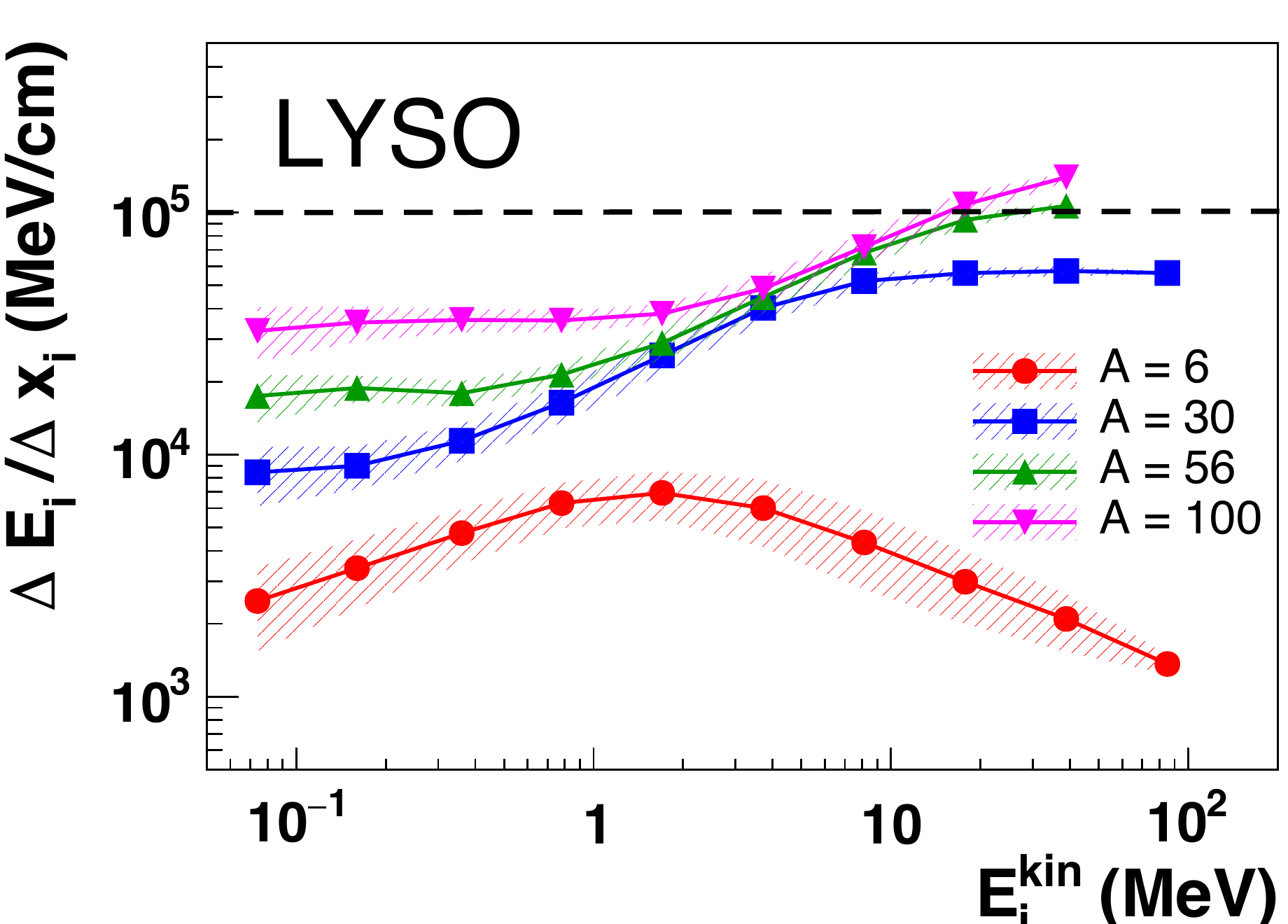}}} &
{\mbox{\includegraphics[viewport = 25 200 555 580, clip, width=46mm]{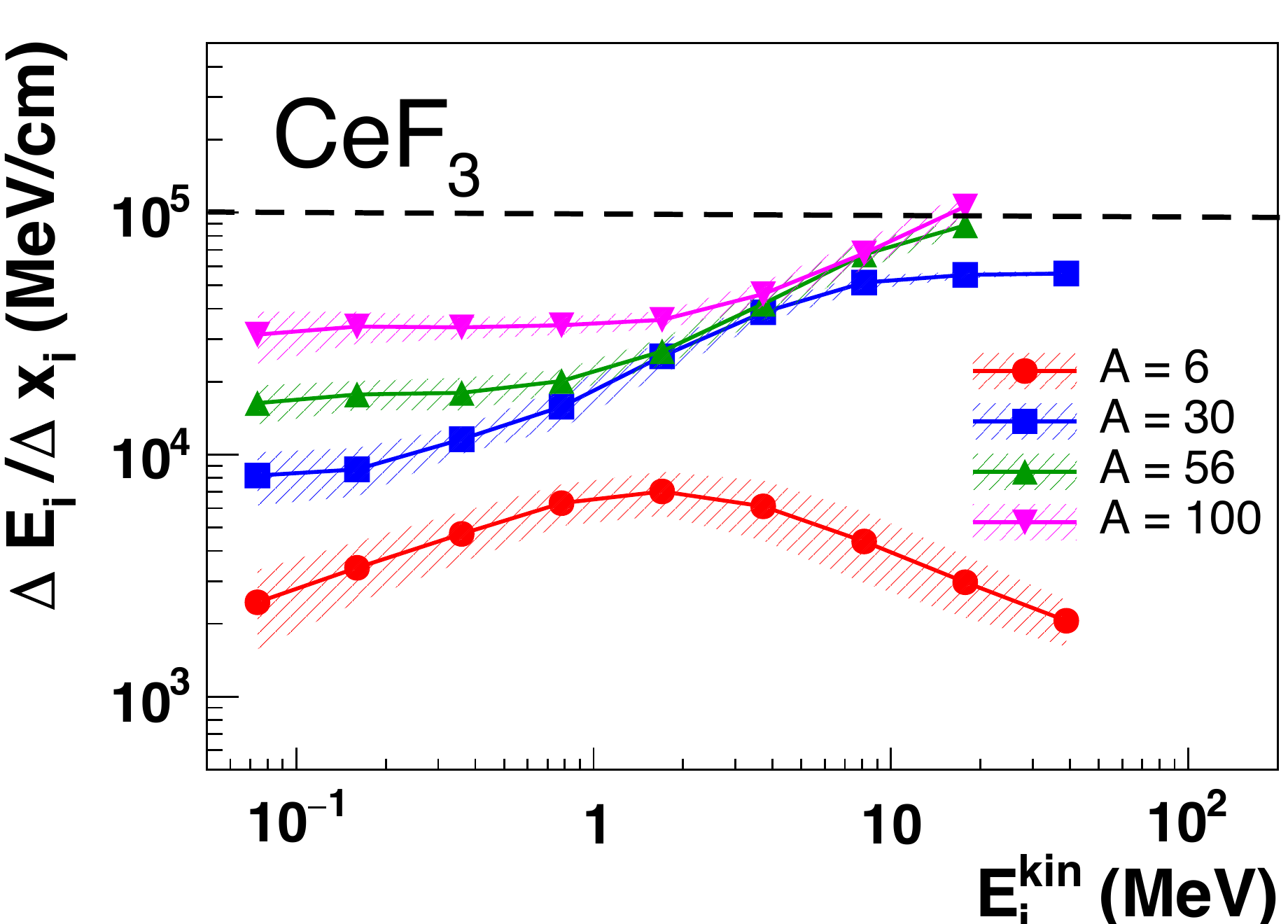}
}}
   \end{tabular}
\end{center}
\caption{Average energy loss for secondaries at every infinitesimal step for PWO (left), LYSO (center) and cerium fluoride (right), as generated by FLUKA, as a function of kinetic energy at the beginning of each step, for given atomic mass numbers (see text for details). The dashed bands indicate the rms of the distributions.}
 \label{f-INF}
\end{figure}
\begin{figure}[b]
\begin{center}
 \begin{tabular}[h]{ccc}
{\mbox{\includegraphics[viewport = 16 10 515 350, clip, width=47mm]{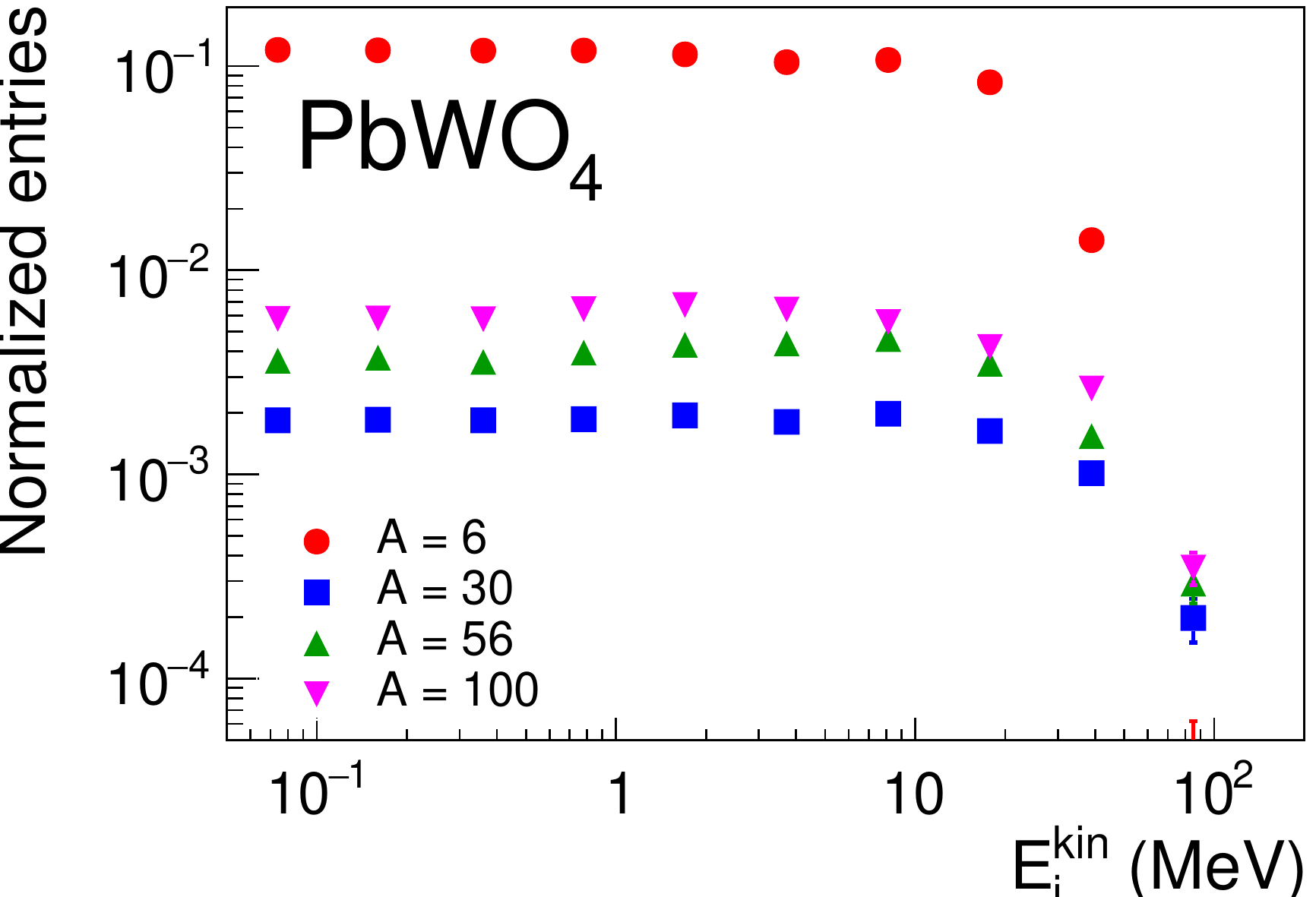}}} &
{\mbox{\includegraphics[viewport = 16 10 515 350, clip, width=47mm]{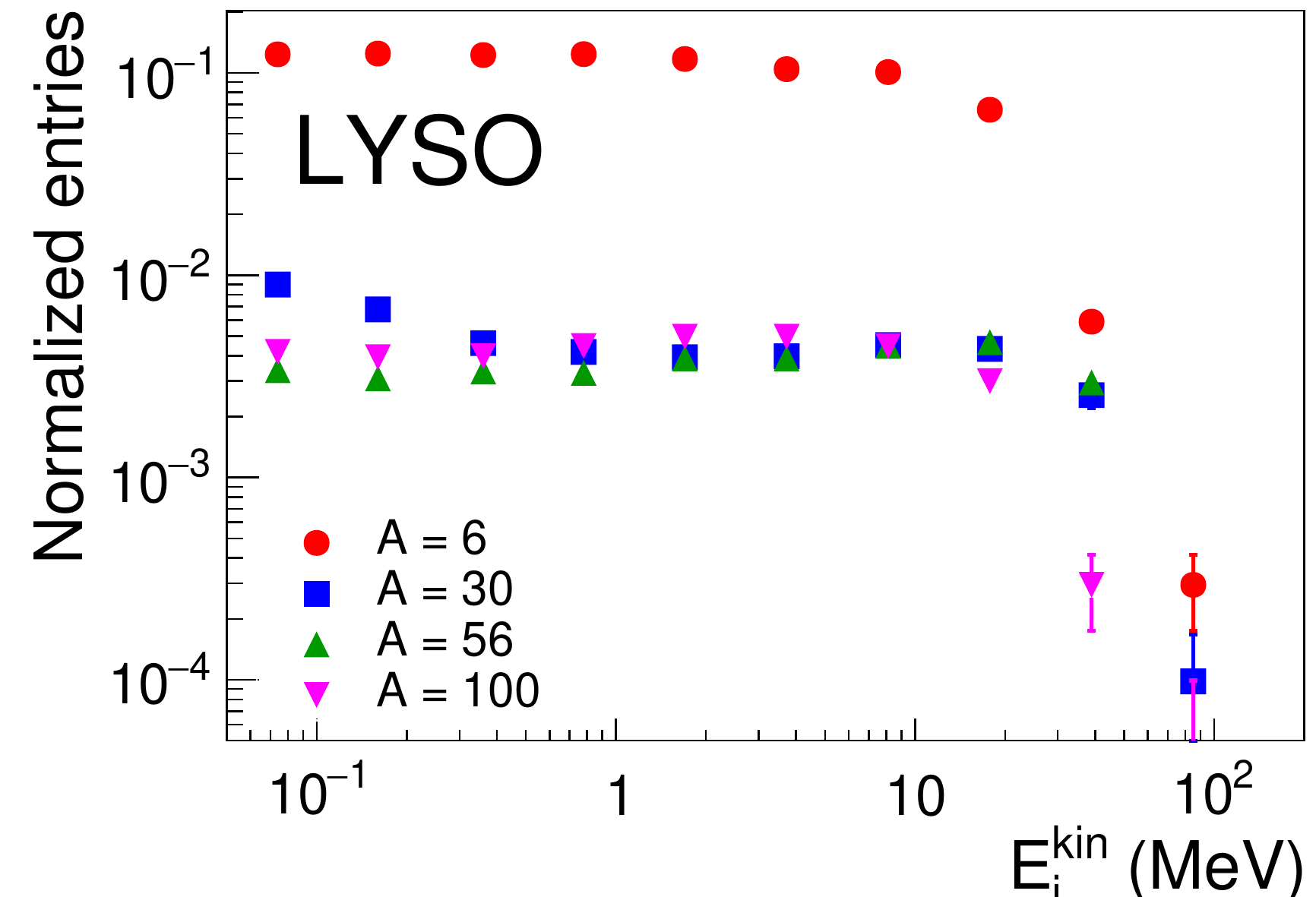}}} &
{\mbox{\includegraphics[viewport = 16 10 515 350, clip, width=47mm]{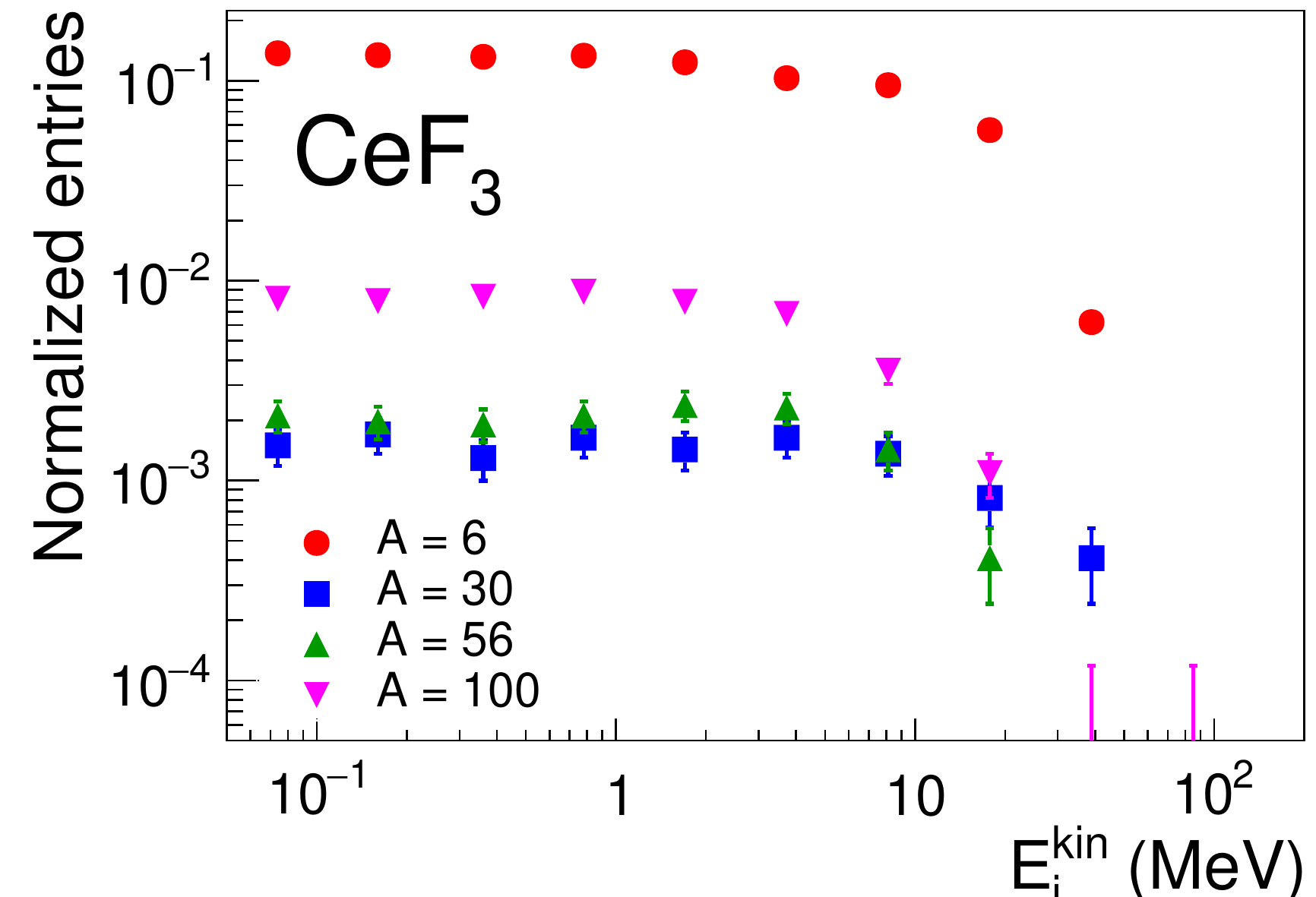}
}}
   \end{tabular}
\end{center}
\caption{Population ratio of infinitesimal steps for secondaries in PWO (left), LYSO (center) and cerium fluoride (right) generated by FLUKA, as a function of kinetic energy at the beginning of each step, for a few given atomic mass numbers (see text for details).}
 \label{f-RAT}
\end{figure}
For each beam proton impact on the crystal, every secondary particle (briefly ``secondaries'') has been generated and transported over infinitesimal steps $\Delta x_i$. At every step $i$, the initial particle kinetic energy $E^{kin}_i$ has been retrieved and the energy loss has been calculated as $\left (\frac{dE}{dx}\right)_i \simeq \left(\frac{\Delta E_i}{\Delta x_i}\right)$ with $\Delta E_i$ the energy lost over the step, and $\Delta x_i$ has been set by FLUKA to allow the fraction of kinetic energy to be lost in a step to be at most 5\%.

Figure~\ref{f-INF} presents the $\left(\frac{\Delta E_i}{\Delta x_i}\right)$ values averaged over all steps, in bins of $E^{kin}_i$ and for given values of particle mass number. 
One observes how energy loss values exceeding
\begin{equation}
dE/dx = 1 \times 10^5 ~\mathrm{MeV/cm}
\label{e-dEdx}
\end{equation}
are reached only in PWO and LYSO, and there just for heavy fragments with mass number $A \gg 30$, in agreement with literature, as already discussed in section~\ref{s-FISS}. For cerium fluoride, no fragments exceed local energy loss values $dE/dx = 1 \times 10^5$ MeV/cm as would be needed for track formation, in agreement with experimental evidence~\cite{r-CEFNIM} of no hadron-specific damage in this material. This is also illustrated in figure \ref{f-RAT}: the fractions of steps by secondaries are plotted, and it is clearly visible where the drop in population occurs for a given A value. For this reason, in the following, studies in cerium fluoride have not been pursued any further, due to the substantiated absence of fission track formation. A first milestone can be considered as reached, however:
the agreement between the microscopic observations from simulations described in this section and the experimental results, of no hadron-specific damage in cerium fluoride, constitute a first validation of the predictive power of the FLUKA simulations.
\begin{figure}[b]
\begin{center}
 \begin{tabular}[h]{cc}
{\mbox{\includegraphics[viewport = 280 200 840 560, clip, width=72mm]{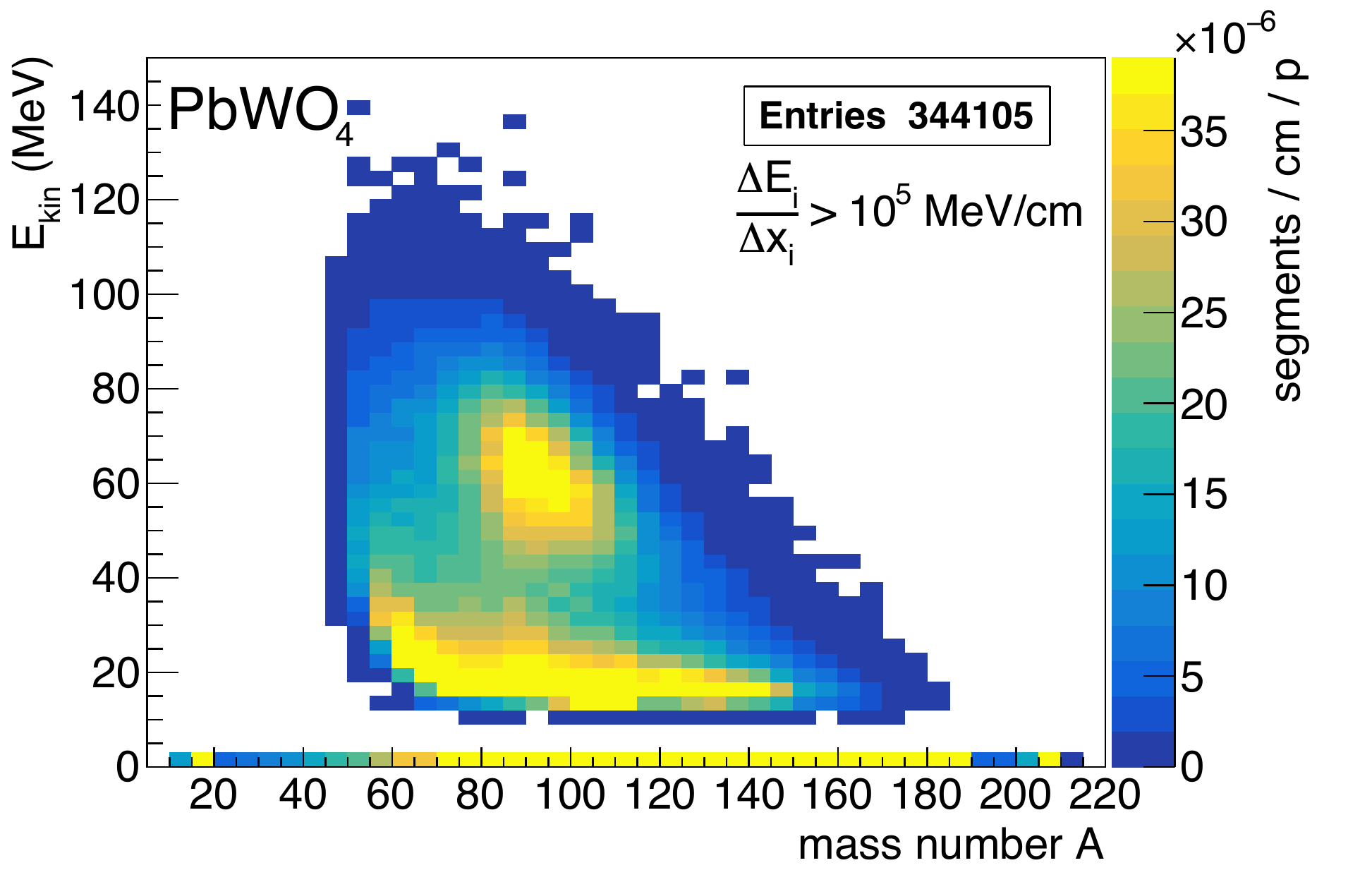}}} &
{\mbox{\includegraphics[viewport = 280 200 840 560, clip, width=72mm]{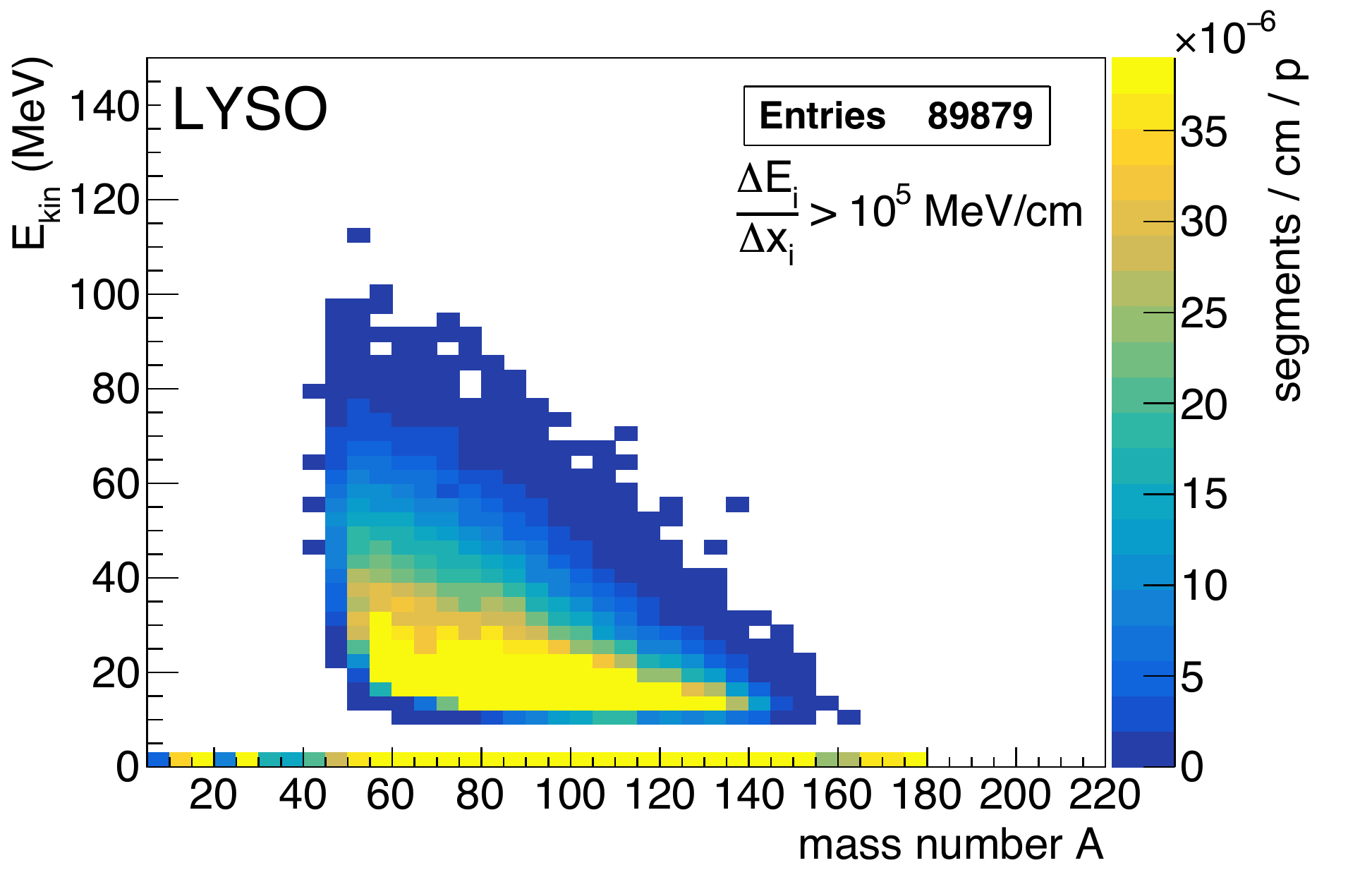}
}}
   \end{tabular}
\end{center}
\caption{In PWO (left) and in LYSO (right), correlation between kinetic energy and mass number for an energy loss requirement of $dE/dx \geq 1 \times 10^5$ MeV/cm.}
 \label{f-SEGnocut}
\end{figure}

\subsection{Simulations of fission track segments}
\label{s-SEG}
\begin{figure}[t]
\begin{center}
 \begin{tabular}[h]{cc}
{\mbox{\includegraphics[viewport = 280 200 840 560, clip, width=72mm]{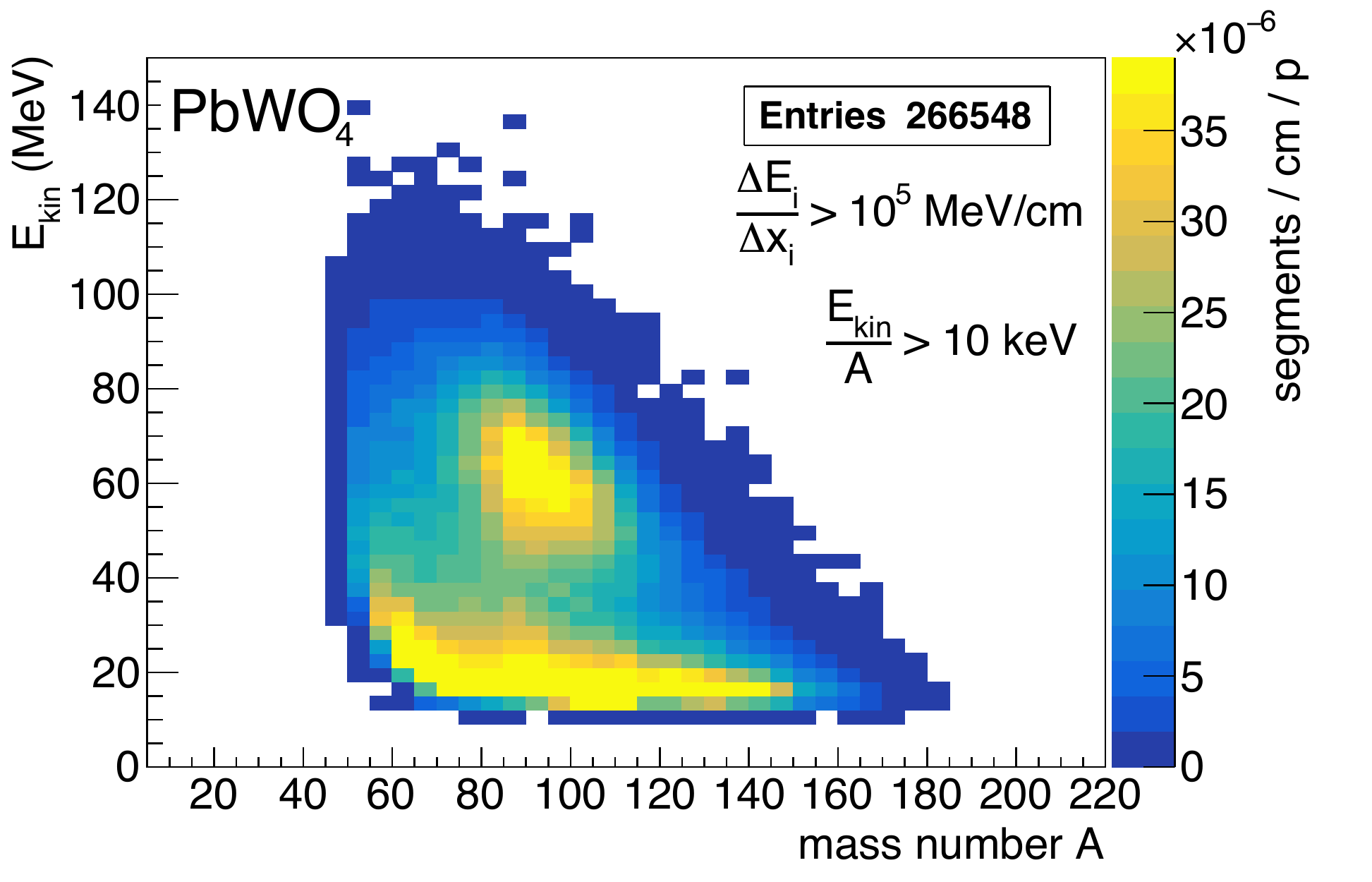}}} &
{\mbox{\includegraphics[viewport = 280 200 840 560, clip, width=72mm]{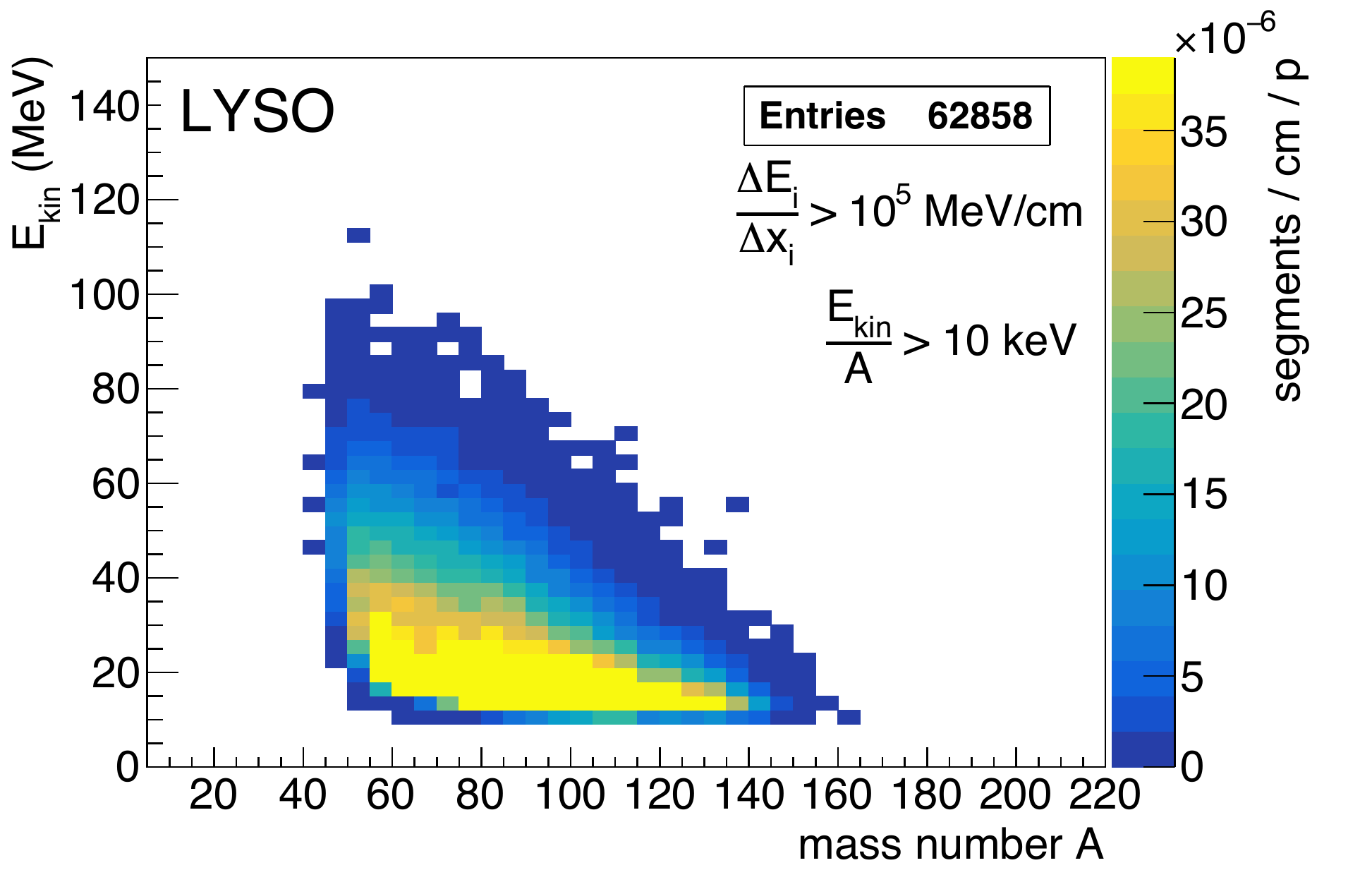}
}}
   \end{tabular}
\end{center}
\caption{In PWO (left) and in LYSO (right), correlation between kinetic energy and mass number for an energy loss requirement of $dE/dx \geq 1 \times 10^5$ MeV/cm and $E^{kin}/A > 10$ keV.}
 \label{f-SEG}
\end{figure}
Fission tracks can be composed of several segments, separated by gaps where the energy loss falls below the required threshold for track formation (see section~\ref{s-FISS}). In the simulations presented here, consecutive steps have been merged into one track segment if the requirement on energy loss in equation~\ref{e-dEdx} is satisfied by both of them. Following this, the length of each segment, the number of segments in a track and the number of segments per incoming proton have been determined.
The distribution of secondary particles kinetic energy values versus mass number is shown in figure \ref{f-SEGnocut}, for a reconstruction requiring an energy loss $dE/dx \geq 1 \times 10^5$ MeV/cm for individual steps to be joined into segments.
The range of mass numbers that give origin to segments in simulation are observed to lie above $A \simeq 40$, in agreement with literature as summarised in section~\ref{s-FISS}. However, there is an exception: a region of $E^{kin} \ll 1$ MeV where all mass number values are populated. These entries disappear though if $E^{kin}/A > 10$ keV is required, as visible in figure \ref{f-SEG}.
The requirement of a minimum threshold in kinetic energy per nucleon finds a justification if one estimates the infinitesimal displacement of such a secondary: the maximum displacement that can occur is $\Delta x_{max} \simeq 10\; $keV$/ (1\times 10^5 $ MeV/cm$) = 10$~\AA, which is of the order of the typical crystal lattice cell size. Such secondaries are therefore fragments that do not have sufficient kinetic energy to move away from their lattice location, and thus cannot possibly produce a track. 
\subsection{From segments to tracks}
\label{s-SEGTRA}
\begin{figure}[t]
\begin{center}
 \begin{tabular}[h]{cc}
{\mbox{\includegraphics[width=70mm]{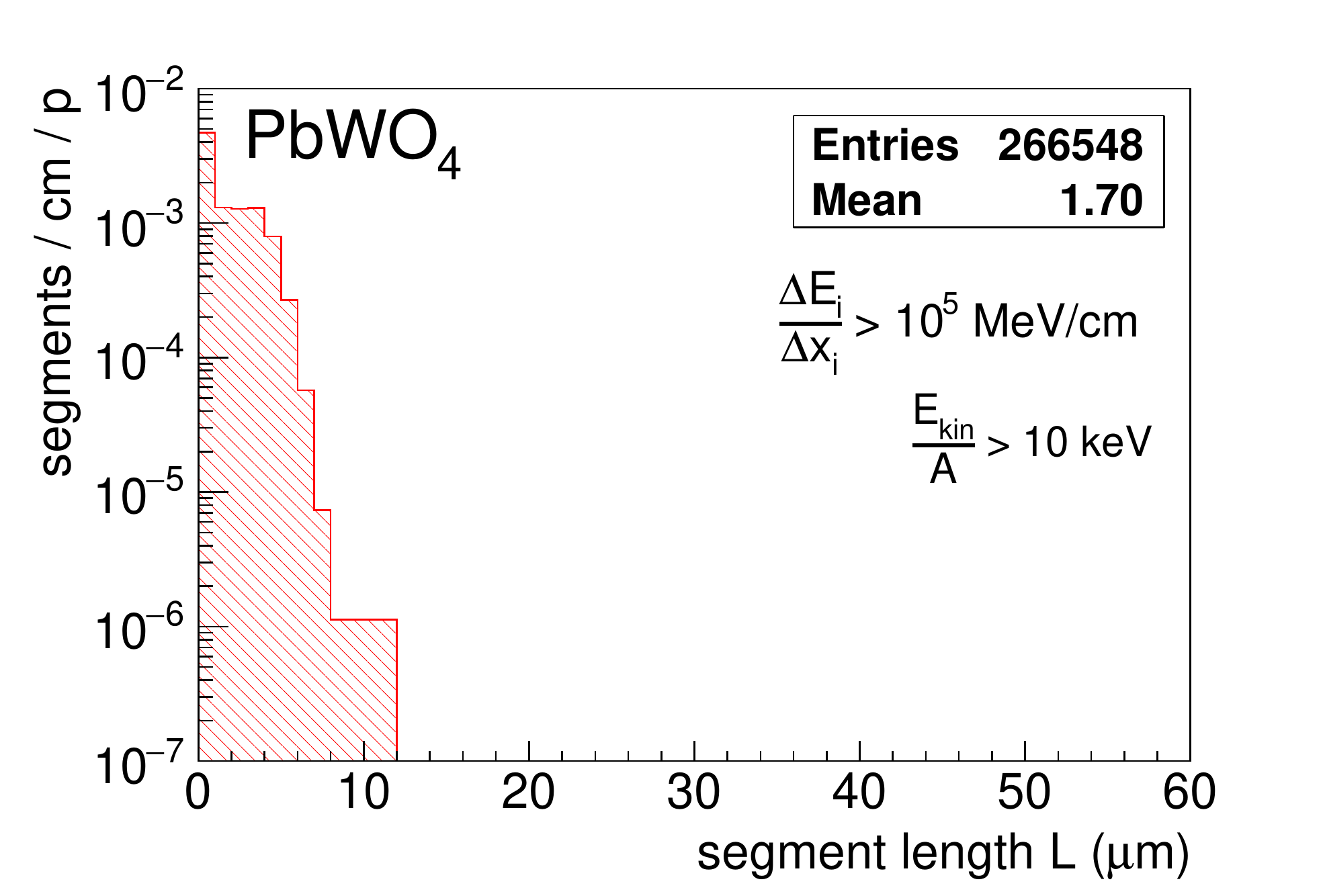}}} &
{\mbox{\includegraphics[width=70mm]{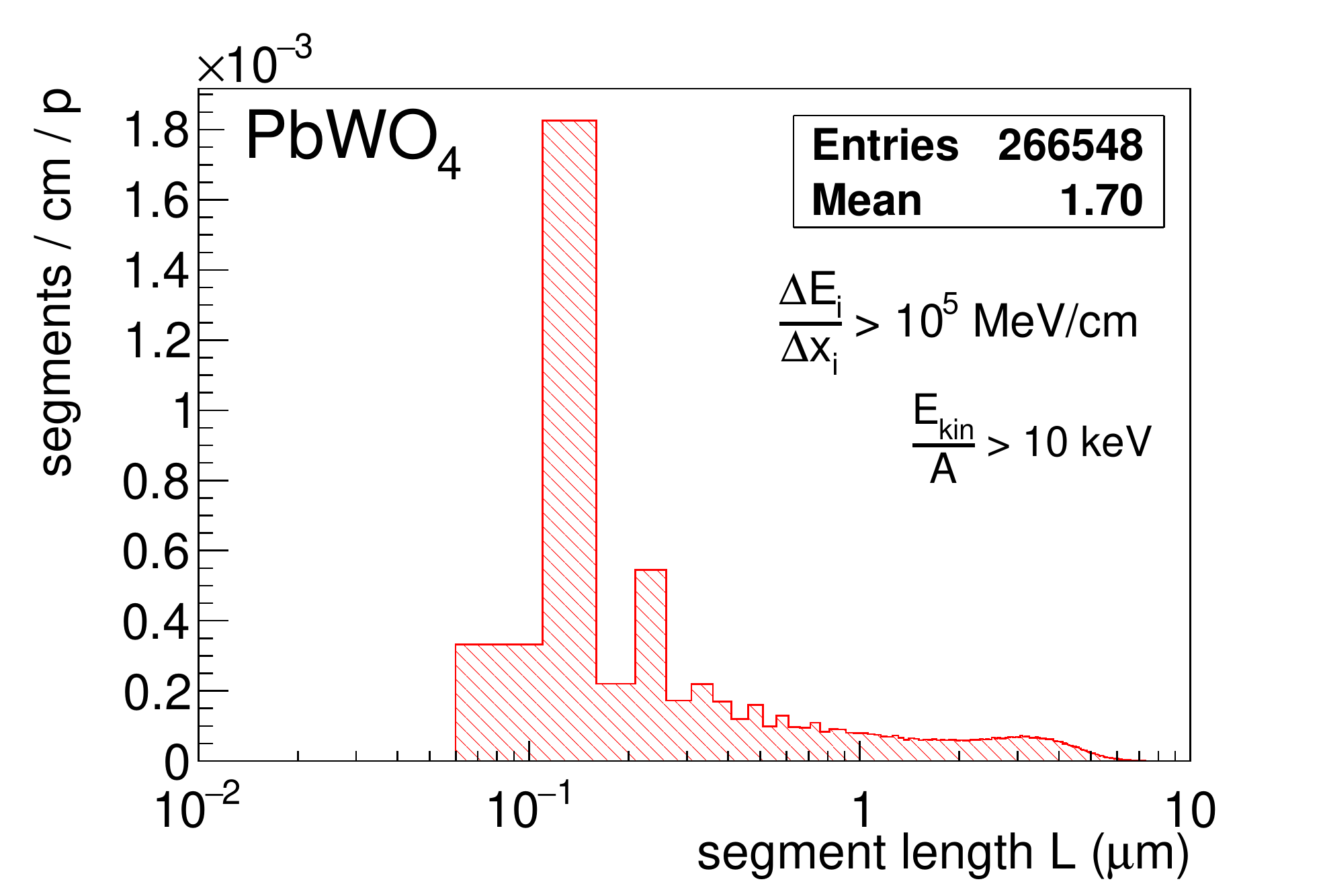}}} \\
{\mbox{\includegraphics[width=70mm]{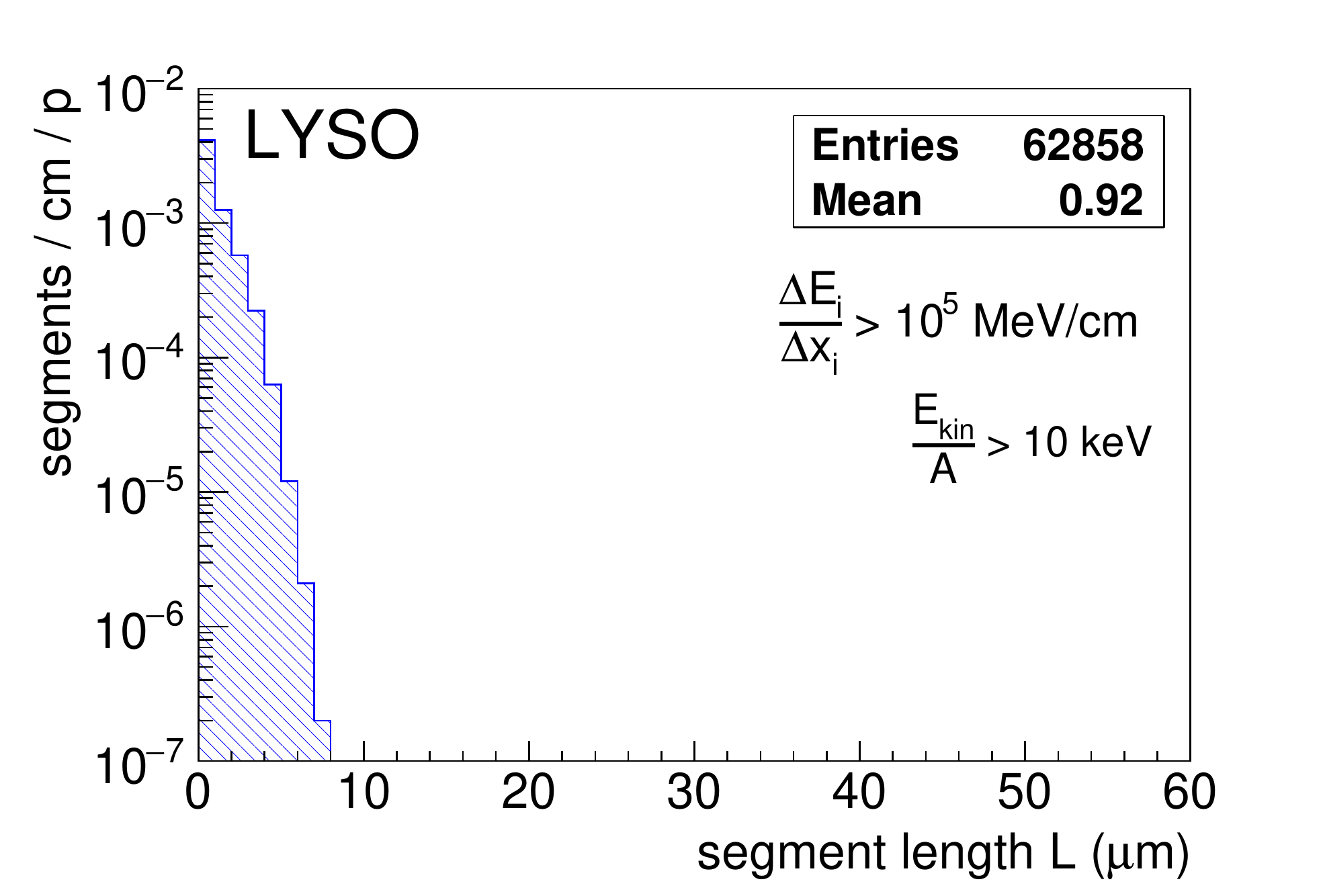}}} &
{\mbox{\includegraphics[width=70mm]{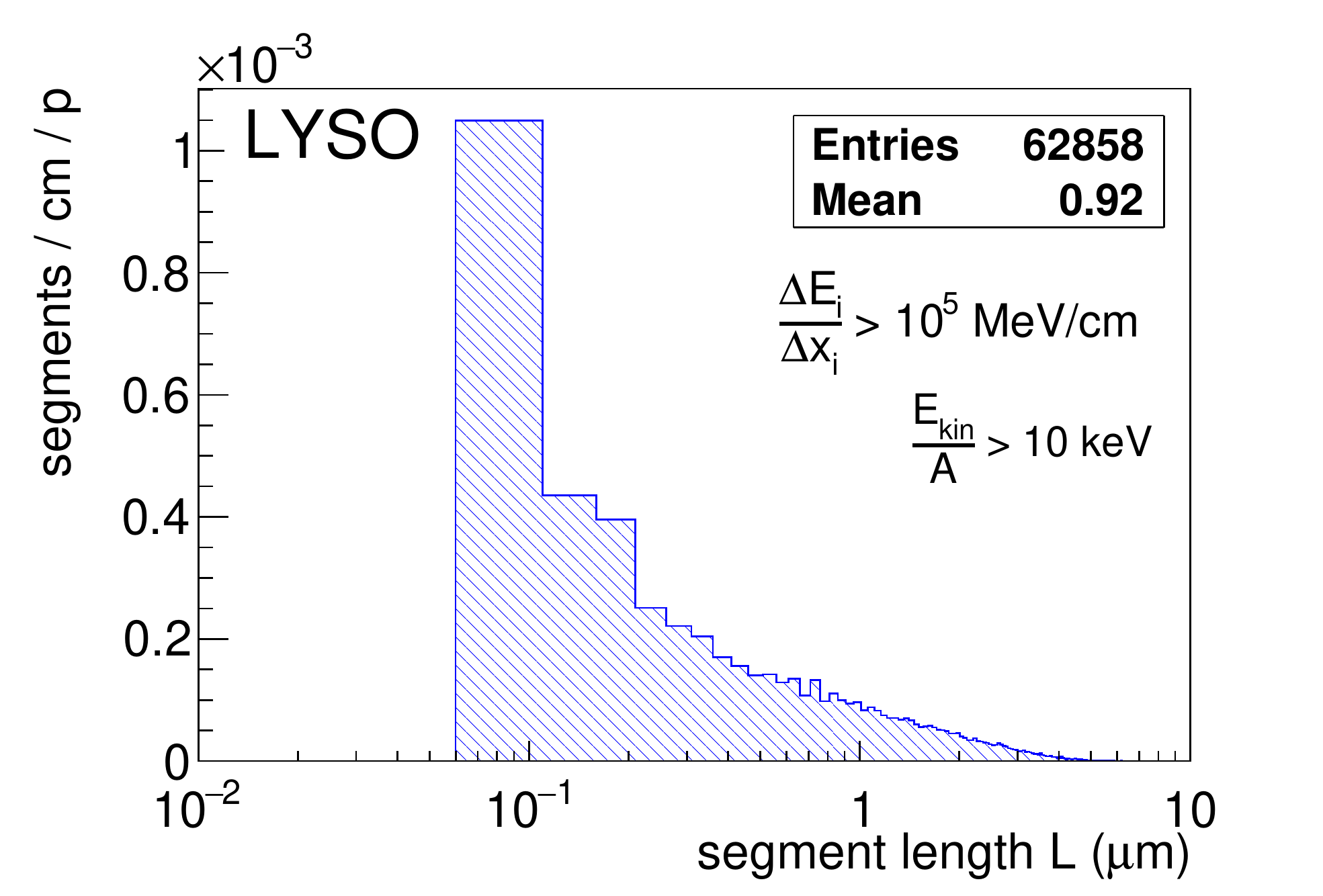}}} \\
   \end{tabular}
\end{center}
\caption{Histograms of simulated segment lengths in PWO (top) and in LYSO (bottom) where a logarithmic scale has been adopted once for the ordinate (left) and once for the abscissa (right).}
 \label{f-LSEGH}
\end{figure}

For the determination of track densities, towards a comparison with observed ones, the full spectrum of simulated track length distributions that are plotted in figure \ref{f-LSEGH} needs to be considered. To allow appreciating more easily the features of the distributions, these have been histogrammed once with a vertical logarithmic scale, and once with a horizontal logarithmic one. One observes that the length of segments from simulations extends over three orders of magnitude, between a few nm and a few $\mu$m.
A comparison can be made with figure 5 from \cite{r-COS}, where two extreme cases are presented: the spectrum of track lengths observed in earth rocks containing radioactive isotopes is limited to short tracks ($< 15 \mu$m), since these are due to fission recoils, while in meteorites a tail of longer tracks can be observed, due to the exposure to heavy, very high-energy cosmic rays, extending to 160 $\mu$m. The distributions in our test (figure \ref{f-LSEGH}, left) have a shape compatible with the observations in terrestrial rocks. This is understandably the case, since the high-energy primary protons in the irradiation tests produce a hadron shower in the crystalline matter, where heavy fission fragments just reach intermediate energies, compared to the two extremes quoted above.
\begin{figure}[b]
\begin{center}
 \begin{tabular}[h]{cc}
{\mbox{\includegraphics[viewport = 290 190 750 560, clip, width=74mm]{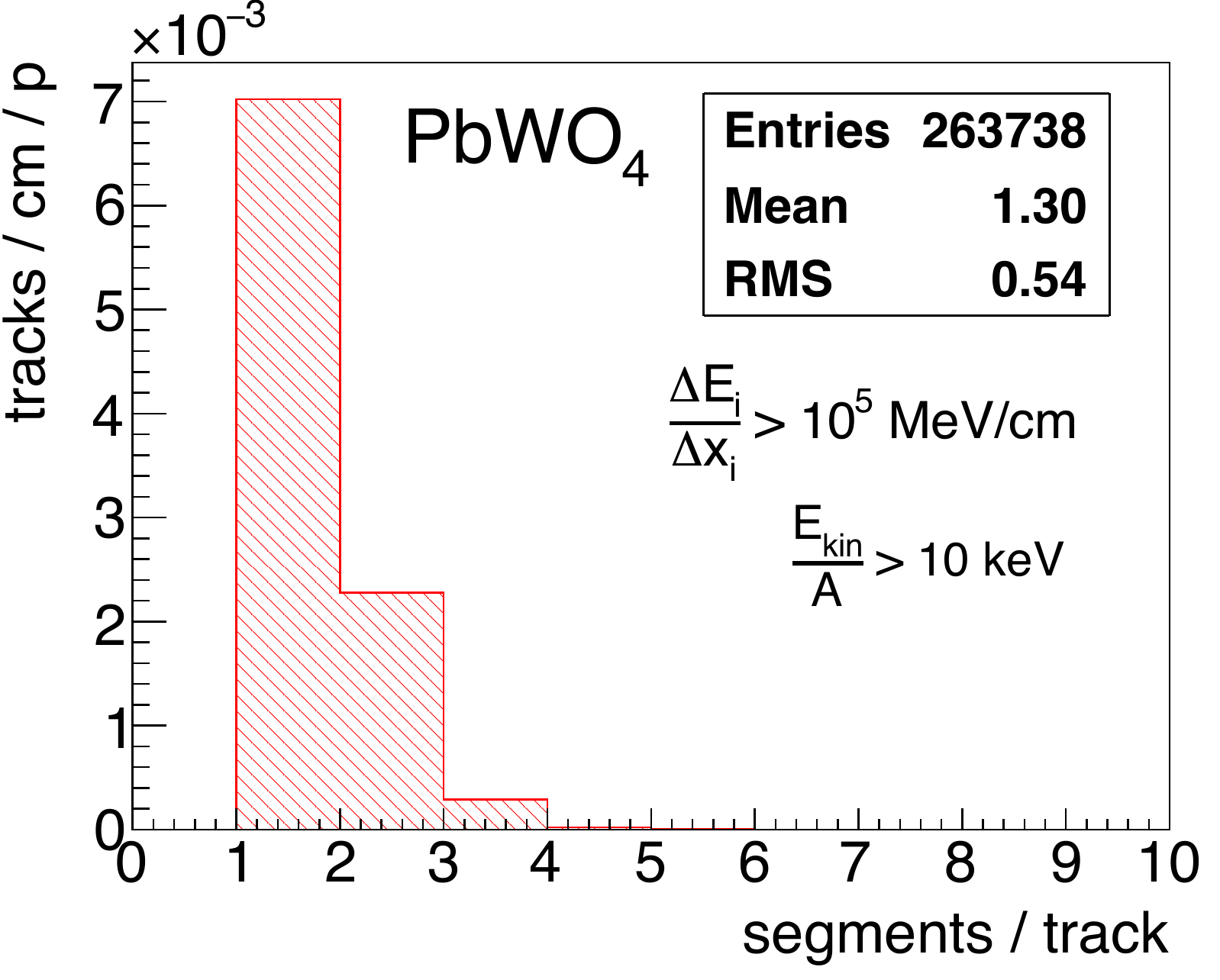}}} &
{\mbox{\includegraphics[viewport = 290 190 750 560, clip, width=74mm]{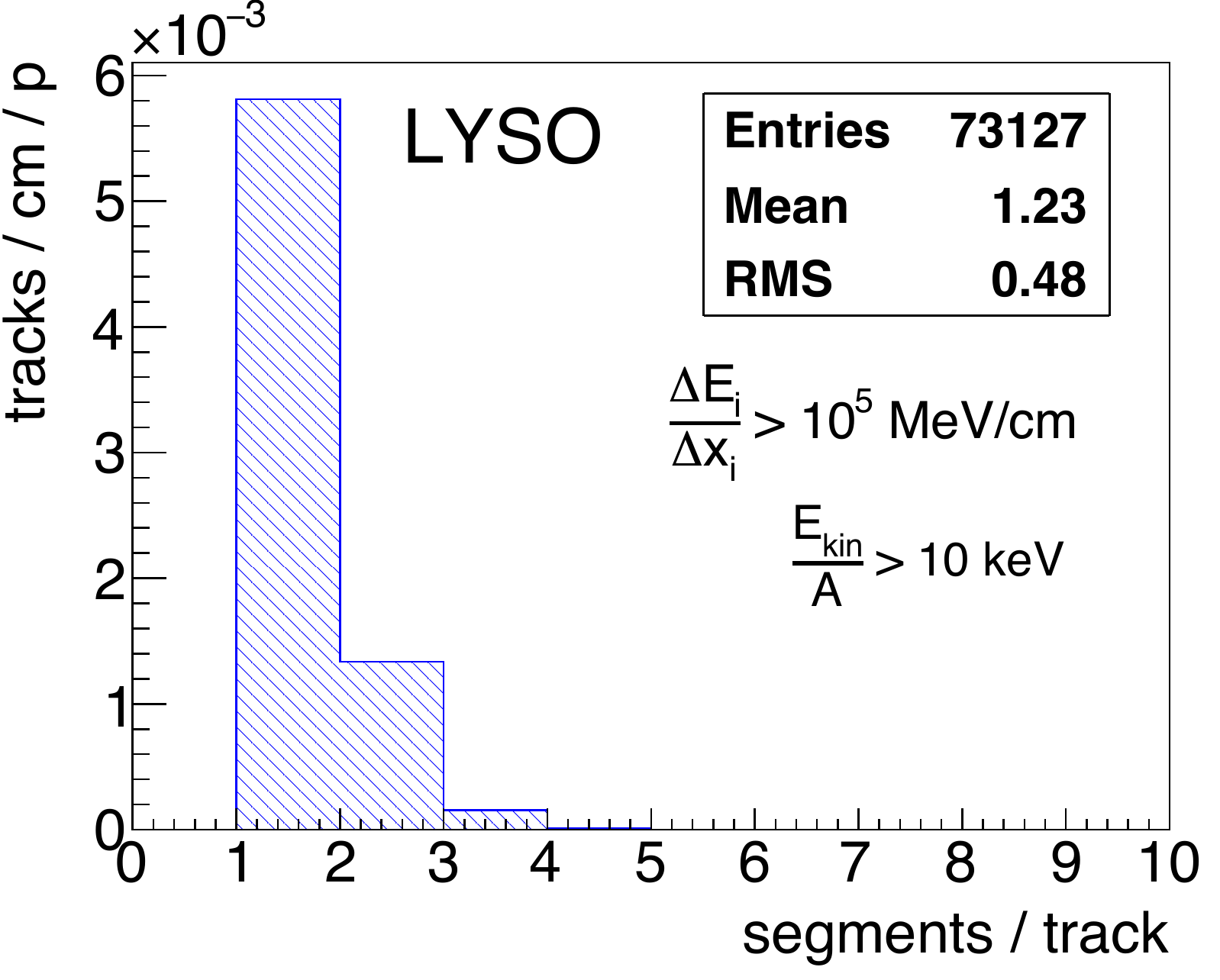}}} 
 \end{tabular}
\end{center}
\caption{Histograms of the number of segments per track from simulations in PWO (left) and in LYSO (right).}
 \label{f-SpT}
\end{figure}

The number of segments belonging to the same track are histogrammed in figure \ref{f-SpT}. While segments in latent tracks are scattering centres for light, in the process of track etching, which is performed for visualisation purposes, those segments get joined, as explained in section~\ref{s-FISS} and in~\cite{r-WAG}. This is evident in the images of etched tracks, as found in figure~\ref{f-PWOTRK} (left) or in \cite{r-FISSNIM} e.g., where no pattern of tracks split into individual segments can be observed. To reach a further validation of the simulations, densities have thus to be compared at the track level. 

From~\cite{r-FISSNIM}, where fission tracks from lead tungstate have been visualised, one obtains an observed track density crossing a surface per proton entering the crystal that amounts to \\$\phi = (1.76\pm 0.14) \times 10^{-6}$ tracks cm$^{-2} /  ($p cm$^{-2})$.
In the FLUKA simulations for lead tungstate, the irradiation has been performed with a proton fluence of $1\times 10^6$ protons on the crystal front face, i.e. $\Phi_p = 2.07\times 10^5$ p cm$^{-2}$. From the number of entries in figure \ref{f-LSEGH} and the lead tungstate crystal volume, the density of segments in the crystal results to be $\rho_{seg} = (1.01\pm 0.25)\times 10^{-2}$ segments cm$^{-3} /  ($p cm$^{-2})$. Here, the statistical uncertainty is negligible, the dominating contribution being due to the dE/dx threshold used. 
From figure \ref{f-SpT} one reads off an average of $1.3$ segments/track
and thus a track density is obtained from FLUKA $\rho_{trk} = (7.8 \pm 2.0)\times 10^{-3}$ tracks cm$^{-3} /  ($p cm$^{-2})$ for PWO.
One can estimate the average track length considering that only tracks with a length of the order of $\left <L\right >$ will be able to traverse a volume of thickness $L$. To get an estimate, we can thus write approximately
\begin{equation}
\rho_{trk} \left <L\right > \simeq \phi
\end{equation}
wherefrom, in our case, we obtain $ \left <L\right > \simeq (2.3\pm 0.6) \mu$m. 

The average length obtained can be compared to the average etched track length visible in figure \ref{f-PWOTRK}, where it becomes evident that the calculated and the observed length agree in order of magnitude, yielding a second, independent validation of the FLUKA simulations.

It might be useful to point out that the track length was inferred from densities, rather than from a direct determination of the simulated track length: this is justified because we want to compare simulation findings with the length of observed etched tracks~\cite{r-FISSNIM}, which tend to be longer than latent ones. It should also be noted that the only tracks directly accessible to simulations are latent ones, and that their length in lead tungstate is found to be on average 2.2~$\mu$m, as can be inferred from figures~\ref{f-LSEGH} and \ref{f-SpT}. 

\subsection{Rayleigh Scattering ratio}
\label{s-RSratio}
After having validated the simulations in two independent ways, confidence was gained that the typical dimensions and densities of segments from simulations can be used to obtain the RS ratio between different crystalline materials. Recalling eq.~\ref{e-RshortS}, we shall call the two intervening fractions
\begin{equation}
R_N = \frac{N_{PWO}}{N_{LYSO}} 
\end{equation}
and 
\begin{equation}
R_d = \frac{d_{PWO}}{d_{LYSO}},
\end{equation}
so that eq.~\ref{e-RshortS} becomes
\begin{equation}
R_{\mu} = \frac{\mu_{PWO}}{\mu_{LYSO}} = 2.46 \times  R_N \times \left( R_d \right)^6.
\label{e-RminS}
\end{equation}
To obtain an estimate of the Rayleigh scattering ratio, one has to take into account some caveats: the RS approximation assumes spheres, while dipole-shaped, randomly oriented tracks are present in calorimeter crystals after exposure to high-energy hadrons. However, for cylindrical scatterers, RS behavior has been observed~\cite{r-MIS}, where either the radius of equivalent-surface spheres or the smallest particle dimension appears to be a good parameter for an estimate~\cite{r-CYL}.

To understand the reliability of simulations, we estimate the effective scatterer dimension for each crystal type individually using the experimental results for the damage amplitudes and the already validated FLUKA segment densities. 

From Eqs.~\ref{e-FRS1} and \ref{e-FRS2}, it follows that 
\begin{equation}
\mu = N \times \sigma_{RS} .
\label{e-musigma}
\end{equation}
This equation, combined with Eq.~\ref{e-sRS}, allows to extract the relevant dimension for Rayleigh scattering. For PWO irradiated with a proton fluence $\Phi_p = 2.07\times 10^5$ p cm$^{-2}$, the density of segments from FLUKA amounts to $N=2100$ segments cm$^{-3}$ and the radiation-induced absorption coefficient is $\mu = 4.57\times 10^{-8}$ m$^{-1}$, as of figure 15 in~\cite{r-LTNIM}, so that the effective scatterer dimension obtained is $d_{PWO} = 32.0$ nm (uncertainties are discussed below). For LYSO, analogously, the density of segments from FLUKA amounts to $N=1006$ segments cm$^{-3}$ for an irradiation with a proton fluence $\Phi_p = 1.6\times 10^5$ p cm$^{-2}$. The measured radiation-induced absorption coefficient is $\mu = 7.8\times 10^{-9}$ m$^{-1}$, as obtained from~\cite{r-LYSONIM} combined with figure 15 in ~\cite{r-LTNIM}. These values lead to an effective scatterer dimension $d_{LYSO} = 31.7$ nm.

These sizes, well compatible between the two crystal types, are nearly an order of magnitude smaller than the scintillation emission wavelength, thus compatible with the requirements of the RS approximation in \ref{e-RSx}. Before considering the uncertainties affecting the estimated dimensions, we compare them to the equivalent-sphere radii that result from purely geometric considerations.

The latent segment lengths obtained from FLUKA are, for PWO, $\left <L\right > = 1700$ nm and,  for LYSO, $\left <L\right > = 920$ nm (figure~\ref{f-LSEGH}). The latent track diameter found in literature is approximately $2.5$ nm for Zircon~\cite{r-YAD}. Using this value yields an estimate for the surface-equivalent sphere radius $R_{PWO} = 46$ nm for PWO and $R_{LYSO} = 34$ nm for LYSO. While this estimate is strongly affected by the uncertainty on the track radius, which is not known for the two crystals considered, it offers a validation of the order of magnitude of the values determined above.

From the estimates based on experimentally validated quantities above, the assumption that $R_d\simeq 1$ appears justified. The RS ratio is then obtained from the segment densities per proton hitting a unit surface derived from the total number of segments (see e.g. figure~\ref{f-LSEGH}), which are, for PWO,
$\rho_{seg}^{PWO} = 1.01\times 10^{-2} \; {\mathrm{segments\; cm}}^{-3} /  ({\mathrm{p\; cm}}^{-2})$
and, for LYSO, 
$
\rho_{seg}^{LYSO} = 0.63\times 10^{-2} \; {\mathrm{segments\; cm}}^{-3} /  ({\mathrm{p\; cm}}^{-2}),$
 with uncertainties discussed below,
 resulting in a RS ratio 
$R_{\mu} = 4.0 $.
\\

Uncertainties on the prediction for $R_{\mu}$ arise from various sources: 
\begin{enumerate}[(a)]
\item the uncertainties on the refractive indices for PWO ($\pm 0.01$~\cite{r-nCHI}) and for LYSO ($\pm 0.005$~\cite{r-nLYSO}) 
contribute with $\Delta R_{\mu, n} = \pm 0.08$;
\item uncertainties in the shape of the emission spectra of LYSO and lead tungstate affect the precision of the factor $\left(\nicefrac{\lambda^{LYSO}}{\lambda^{PWO}}\right)^4$ determination and contribute 
with $\Delta R_{\mu, em} = \pm 0.2$;
\item the uncertainties on the two ratios of refractive indices between metamict and crystalline state for PWO and for LYSO, $\Delta m_{PWO}=\Delta m_{LYSO}= \pm 0.01$, cause a significant 
contribution $\Delta R_{\mu, m} = \pm 1.60$;
\item The uncertainty on the applicable dE/dx threshold affects the segment densities for PWO and LYSO in a correlated way, resulting in a contribution $\Delta R_{\mu ,dEdx1} = \pm 0.6$;
\item A common dE/dx threshold  has been used for both, PWO and LYSO. Allowing for a shift between the two threshold values causes an uncertainty $\Delta R_{\mu ,dEdx2} = \pm 0.8$;
\item The uncertainties on $d$ for PWO and for LYSO from segment densities
contribute with $\Delta R_{\mu, d}=\pm 0.24$; 
\item The total statistical uncertainty of  $N_{PWO}$ and $N_{LYSO}$
contributes with $\Delta R_{\mu, N}=\pm 0.04$;
\end{enumerate}
The total uncertainty thus amounts to  $\Delta R_{\mu}= \pm 1.9$ and the FLUKA prediction is given by
\begin{equation}
R_{\mu} = 4.0 \pm 1.9.
\end{equation}

This result has to be compared with the measured ratio $R_{\mu}^{meas}= 4.5 \pm 0.2$: the Rayleigh scattering amplitude ratio between PWO and LYSO obtained from FLUKA simulations is compatible, within the uncertainties, with the measured one. 
However, differences in scatterer dimensions could lead to significantly different $R_{\mu}$ values, due to the presence of a $6^{th}$ power dependence.

This last validation, of the RS ratio, along with those described in section~\ref{s-INF} for the fulfilment of track-formation requirements and in section~\ref{s-SEG} for the segment lengths, corroborates the possibility of estimating, through simulations, the order of magnitude of relative damage amplitudes to be expected from hadrons in different inorganic crystals.
\section{Conclusions}
\label{s-CON}
A simulation study has been performed of long, inorganic crystals irradiated by 24 GeV protons producing hadron showers in the materials, and several observable quantities have been compared with measured ones. 
The simulations confirm the qualitative understanding that has been reached of the hadron-specific damage component in the studied inorganic scintillating crystals. In particular, FLUKA simulations yield no heavy, highly ionising fragments in cerium fluoride, as would be needed for track creation, in agreement with the observed absence of a hadron-specific damage component~\cite{r-CEFNIM}. 
Instead, the simulations yield heavy, highly ionising fragments in PWO and LYSO, as needed for track creation, in agreement with the observed hadron specific damage~\cite{r-LTNIM,r-LYSONIM}. In lead tungstate, the simulations yield track densities that are compatible with experimentally observed ones
~\cite{r-FISSNIM}. Further, the simulated spectrum of segment lengths is in qualitative agreement with experimental data~\cite{r-FLEME}. Finally,  FLUKA simulations yield a Rayleigh Scattering amplitude ratio between PWO and LYSO that is consistent, within the uncertainties, with the measured one~\cite{r-LYSONIM}. All these different points of validation confirm the reliability of FLUKA simulations, deployed at a microscopic level according to criteria that were established experimentally~\cite{r-FLE}, in estimating the order of magnitude of damage densities  and amplitudes that should be expected from hadrons in different inorganic crystals, enabling thus an informed choice of materials in calorimeter design.
\acknowledgments
The many helpful suggestions received from the FLUKA community through its discussion forum are warmly acknowledged, as well as Sophie Mallows' help in setting up the simulation framework.

\end{document}